\documentclass{llncs}
\usepackage{amssymb}
\usepackage{latexsym}
\usepackage{enumerate}
\usepackage{amsmath}
\def\adimH{\mathsf{K}} 
\def\aNumbers{\mathcal{N}} 
\def\bmat{\begin{pmatrix}}
\def\emat{\end{pmatrix}}
\def\ExpG{\mathit{E}}
\def\C{\mathbb{C}} 
\def\Clang{\textit{C~}}
\def\classN{\mathsf{m}} 
\def\e{\mathrm{e}}
\def\GAP{\textit{GAP}}
\def\Hspace{\mathcal{H}} 
\def\id{\mathbf{1}} 
\def\idmat{\mathrm{I}} 
\def\ig{\gamma} 
\def\iG{\Gamma} 
\def\iGN{{\cabs{\iG}}} 
\def\iGX{\iG^{\X}} 
\def\ls{\sigma} 
\def\lS{\Sigma} 
\def\lSN{{\cabs{\lS}}} 
\def\lSX{\lS^{\X}} 
\def\natmod{\mathsf{H}} 
\def\N{\mathbb{N}}
\def\NF{\mathcal{F}}

\def\period{\mathcal{P}} 
\def\Prob{\mathrm{\mathbf{P}}} 
\def\Q{\mathbb{Q}} 
\def\regrep{\mathrm{P}} 

\def\repirr{D} 
\def\repq{\mathrm{U}} 
\def\runisymb{\mathsf{r}} 
\def\sg{\mathsf{f}} 
\def\sG{\mathsf{F}} 
\def\sGN{{\cabs{\sG}}} 
\def\tin{0} 
\def\tfin{T} 
\def\Time{\mathcal{T}} 
\def\TprimG{\mathsf{T'}} 
\def\transmatr{\mathrm{T}} 
\def\transmatrprim{\mathrm{T'}} 
\def\wg{\mathsf{g}} 
\def\wG{\mathsf{G}} 
\def\wGN{\mathsf{M}} 
\def\ws{\omega} 
\def\wS{\Omega} 
\def\wSN{\mathsf{N}} 
\def\x{\mathsf{x}} 
\def\X{\mathsf{X}} 
\def\XN{{\cabs{{\X}}}} 
\def\Z{\mathbb{Z}}
\newcommand{\AltG}[1]{\mathsf{A}_{#1}} 
\newcommand{\Aut}[1]{\mathrm{Aut}\vect{#1}} 
\newcommand{\barket}[1]{\left|#1\right\rangle} 
\newcommand{\braket}[1]{\left\langle#1\right\rangle} 
\newcommand{\cabs}[1]{\left|#1\right|} 
\newcommand{\cconj}[1]{\overline{#1}} 
\newcommand{\class}[1]{K_{#1}} 
\newcommand{\CyclG}[1]{\Z_{#1}} 
\newcommand{\DihG}[1]{\mathsf{D}_{#1}} 
\newcommand{\Grassnorm}[1]{\left\|#1\right\|} 
\newcommand{\fra}[2]{\frac{\textstyle{#1}}{\textstyle{#2}}}
\newcommand{\inner}[2]{\left\langle#1\mid#2\right\rangle} 
\newcommand{\innerstandard}[2]{\left(#1\mid#2\right)} 
\newcommand{\invarL}[2]{\mathrm{L}_{#1}\vect{#2}} 
\newcommand{\invarQ}[3]{\mathrm{Q}_{#1}\vect{#2,#3}} 
\newcommand{\Math}[1]{$#1$} 
\newcommand{\Mathh}[1]{$$#1$$} 
\newcommand{\Mone}[1]{\bmat#1\emat} 
\newcommand{\Mthree}[9]{\bmat#1&#2&#3\\
 #4&#5&#6\\
 #7&#8&#9\emat} 
\newcommand{\Mtwo}[4]{\bmat#1&#2\\#3&#4\emat} 
\newcommand{\ordset}[1]{\left[#1\right]} 
\newcommand{\Perm}[1]{\mathrm{Sym}\left(#1\right)} 
\newcommand{\ProbBorn}[2]{\Prob\!\vect{#1,#2}} 
\newcommand{\QuatG}[1]{\mathsf{Q}_{#1}} 
\newcommand{\runi}[1]{\runisymb_{#1}} 
\newcommand{\set}[1]{\left\{#1\right\}} 
\newcommand{\SL}[2]{\mathsf{SL}\vect{#1,#2}} 
\newcommand{\SU}[1]{\mathsf{SU}\vect{#1}} 
\newcommand{\SymG}[1]{\mathsf{S}_{#1}} 
\newcommand{\U}[1]{\mathsf{U}\vect{#1}} 
\newcommand{\vect}[1]{\left(#1\right)} 
\newcommand{\Vthree}[3]{\bmat#1\\#2\\#3\emat} 
\newcommand{\Vtwo}[2]{\bmat#1\\#2\emat} 
\begin{document}
\title{Computations in Finite Groups and Quantum Physics}
\titlerunning{Finite Groups and Quantum Physics}
\author{Vladimir V. Kornyak}
\institute{Laboratory of Information Technologies \\
           Joint Institute for Nuclear Research \\
           141980 Dubna, Russia \\
           \email{kornyak@jinr.ru}}
\authorrunning{Vladimir V. Kornyak}
\maketitle
\begin{abstract}
Mathematical core of quantum mechanics is the theory of unitary representations
of symmetries of physical systems.
We argue that quantum behavior is a natural result of extraction of 
``observable'' information about systems containing
``unobservable'' elements in their descriptions. Since our aim is
physics where the choice between finite and infinite descriptions
can not have any empirical consequences, we consider the problem
in the finite background. Besides, there are many indications from
observations --- from the lepton mixing data, for example --- that
finite groups underly phenomena in particle physics at the deep
level. The ``finite'' approach allows to reduce any quantum
dynamics to the simple permutation dynamics and, thus, to express
quantum observables in terms of permutation invariants of symmetry
groups and their integer characteristics such as sizes of
conjugate classes, sizes of group orbits, class coefficients, and
dimensions of representations. Our study has been accompanied by
computations with finite groups, their representations and
invariants. We have used both our \Clang implementation of
algorithms for working with groups and computer algebra system
\GAP.
\end{abstract}
\section{Introduction}
Symmetry is the leading mathematical principle in quantum mechanics:
only systems containing indistinguishable particles demonstrate quantum
behavior --- any violation of identity of particles destroys quantum interferences.
\par
Mathematical description of any system uses arbitrarily chosen marks for registration and
identification elements of the system. Elements of systems with symmetries are decomposed
into ``homogeneous'' sets --- group orbits. Only such relations and statements (they are
called \emph{invariants}) have objective meaning as are not dependent on relabeling
elements lying on the same group orbit. An example of such invariant is the number
of elements of a group orbit. To fix an element of a group orbit
is possible  only with respect to some additional system which appears as
``\emph{coordinate system}'', or ``\emph{observer}'', or ``\emph{measuring device}''.
For example, no objective meaning can be attached to electric potentials
\Math{\varphi} and \Math{\psi} or to points of space, denoted (marked)
as vectors \Math{\mathbf{a}} and \Math{\mathbf{b}}. But the combinations denoted as
\Math{\psi-\varphi} or \Math{\mathbf{b}-\mathbf{a}} (in more general group notation
\Math{\varphi^{-1}\psi} and \Math{\mathbf{a^{-1}b}})
are meaningful. These are examples of typical situations where observable objects or
relations are group invariants depending on pairs of elements related to observed
system and  to observer.
\par
The question of ``whether the real world is discrete or continuous'' or even ``finite
or infinite'' is entirely \emph{metaphysical}, since neither empirical observations nor
logical arguments can validate one of the two adoptions --- this is a matter of belief or
taste. Since the choice between finite (discrete) and infinite
(continuous) descriptions can not have any empirical consequences
--- ``{physics is independent of metaphysics}'' --- we can boldly take advantage
of  ``finite'' consideration without any risk to destroy the physical content of a problem.
\par
In this paper, we consider \emph{finite quantum mechanics} from
constructive, algorithmic point of view. Using the fact that
\emph{any representation} of finite group can be embedded into a
permutation representation, we show that any quantum dynamics can
be reduced to \emph{permutations}, and quantum observables can be
expressed in terms of \emph{permutation invariants}. Note that the
interpretational issues like ``\emph{wavefunction collapse}'',
``\emph{many-worlds}'', ``\emph{many-minds}'' etc. disappear in
the finite background. We discuss also experimental evidences of
fundamental role of finite symmetry groups in particle physics.
\section{Dynamical Systems and Quantum Evolution}
Let us consider  \textbf{dynamical system} with the finite set of (classical) \textbf{states}
\Math{\wS=\set{\ws_1,\ldots,\ws_{\wSN}}} in the \textbf{discrete time} \Math{t\in\Time},
where \Math{\Time=\Z} or \Math{\Time=\ordset{\tin,1,\ldots,\tfin}}.
We assume that a finite \textbf{symmetry group}
 \Math{\wG=\set{\wg_1,\ldots,\wg_{\wGN}}\leq\Perm{\wS}}
acts on the set of states.
\par
{\textbf{Classical evolution}} (trajectory) of the dynamical
system is a sequence of states evolving in time
\Math{\ldots,\ws_{t-1},\ws_{t},\ws_{t+1},\ldots\in\wS^\Time}.
\par
For reasons that will be clear later, we define \textbf{quantum evolution} as
a sequence of permutations \Math{\ldots{}p_{t-1},p_{t},p_{t+1}\ldots\in\wG^\Time,}
\Math{p_{t}\in\wG}.
\par
In most physical problems, the whole set of states \Math{\wS} has a special structure
of a set of functions \Math{\wS=\lSX} on some \textbf{space} \Math{\X} with
values in some set of \textbf{local states} \Math{\lS}.
In dynamical systems with such structure of the set of states nontrivial
\emph{gauge structures}
--- used in physical theories for description of forces --- arise naturally.
We assume that the space is a finite set \Math{\X=\set{\x_1,\ldots, \x_\XN}}
possessing nontrivial group of \textbf{space symmetries}
\Math{\sG=\set{\sg_1,\ldots,\sg_\sGN}\leq\Perm{\X}}.
The local states form a finite set \Math{\lS=\set{\ls_1,\ldots,\ls_\lSN}}
provided with the group of \textbf{internal symmetries}
\Math{\iG=\set{\ig_1,\ldots, \ig_\iGN}\leq\Perm{\lS}}.
To combine the space \Math{\sG}
and internal \Math{\iG} groups into the symmetry group \Math{\wG} of the whole
set of states \Math{\wS=\lSX} we use the following equivalence class of
\emph{split extensions}
\begin{equation}
\id\rightarrow\iGX\rightarrow\wG\rightarrow\sG\rightarrow\id,
\label{extensionEn}
\end{equation}
where \Math{\iGX} is the group of \Math{\iG}-valued functions on the space \Math{\X}.
This is a natural generalization of constructions used in physical theories.
Explicit formulas for group operations in \Math{\wG} expressed in terms of operations
in \Math{\sG} and \Math{\iG} are given in \cite{KornyakJMS10En,KornyakNS10}
--- we do not need them here.
\par
The most popular and intuitive approach to quantization
---  particularly well suited for dynamical systems with space  ---
is Feynman's path integral:
the amplitude of quantum transition from initial
to final state is computed by summing up the amplitudes along all possible classical
trajectories connecting these states.
As is well known, Feynman's approach is equivalent to the traditional matrix formulation
of quantum mechanics where the evolution of a system
from an initial to a final state is described by an \textbf{evolution matrix} \Math{U}:
\Math{\barket{\psi_{\tin}}\rightarrow\barket{\psi_{\tfin}}=U\barket{\psi_{\tin}}}.
The evolution matrix of a quantum dynamical system can be represented as the product
of matrices corresponding to elementary time steps:
\Math{U=U_{\tfin\leftarrow\tfin-1}\cdots{}U_{t\leftarrow{}t-1}\cdots{}U_{1\leftarrow0}.}
In fact, it can be shown by straightforward examination that Feynman's quantization
rules --- ``multiply subsequent events'' and
``sum up alternative histories'' --- is simply a rephrasing of the matrix
 multiplication rule. For the sake of uniformity of consideration
we adopt the evolution matrix approach throughout this paper.
\par
Quantum mechanical evolution matrices are unitary operators acting
in Hil\-bert spaces of (quantum) \emph{state vectors}
(called also ``\emph{wave functions}'', ``\emph{amplitudes}'' etc.).
\emph{Quantum mechanical particles} are associated with unitary representations
of certain groups. These representations are called ``\emph{singlets}'',
``\emph{doublets}'', and so on, in accordance with their dimensions.
Multidimensional representations describe the \emph{spin}.
A \emph{quantum mechanical experiment} is reduced to comparison of the system
state vector \Math{\psi} with some sample state vector \Math{\phi}
provided by a ``\emph{measuring apparatus}''.
According to the Born rule, the probability to observe the coincidence of
the states is equal to \Math{{\cabs{\inner{\phi}{\psi}}^2}/
\vect{\inner{\phi}{\phi}\inner{\psi}{\psi}}}.
\section{Groups, Numbers and Representations}
All transitive actions of a finite group \Math{\wG=\set{\wg_1,\!\ldots\!,\wg_{\wGN}}}
on finite sets \Math{\Omega=\set{\omega_1,\!\ldots\!,\omega_\wSN}}
can easily be described \cite{HallEn}.
Any such set is in one-to-one correspondence with  \emph{right}
\Math{H\backslash\wG} (or \emph{left} \Math{\wG/H})
\emph{cosets} of some subgroup \Math{H\leq\wG}.
The set \Math{\Omega} is called a \emph{homogeneous space} of the group \Math{\wG}
(\Math{\wG}-\emph{space}).
Action of \Math{\wG} on \Math{\Omega} is \emph{faithful}, if the subgroup \Math{H}
does not contain normal subgroups of \Math{\wG}. We can write action in the form of
permutations
\Mathh{
\pi(g)=\dbinom{\omega_i}{\omega_ig}\sim\dbinom{Ha}{Hag},
\hspace*{20pt}g,a\in{}\wG,~~~i=1,\ldots,\wSN.
}
\par
Maximal transitive set \Math{\Omega} is the set of all elements of the group \Math{\wG}
itself, i.e., the set of cosets of the trivial subgroup \Math{H=\set{\id}}.
The corresponding action is called \emph{regular} and can be represented by the permutations
\begin{equation}
    \Pi(g)=\dbinom{\wg_i}{\wg_ig},~~~~i=1,\ldots,\wGN.
    \label{regpermEn}
\end{equation}
\par
To introduce a ``quantitative'' (``statistical'') description, let us assign
to the elements of the set \Math{\Omega} numerical ``weights'' from some suitable
\emph{number system} \Math{\aNumbers} containing at least \emph{zero} and \emph{unity}.
This allows to rewrite permutations by matrices ---
this is called \emph{permutation representation}:
\begin{equation}
\pi(g)\rightarrow\rho(g)=
\Mone{\rho(g)_{ij}},\text{~~ where~~} \rho(g)_{ij}=\delta_{\omega_ig,\omega_j};
~~ i,j=1,\ldots,\wSN.
\label{permrepEn}
\end{equation}
Here \Math{\delta_{\alpha,\beta}} is the Kronecker delta on
\Math{\Omega}.
\par
The \emph{cycle type} of a permutation is  array of multiplicities
of lengths of cycles in decomposition of the permutation into disjoint
cycles. The cycle type is usually denoted by
\Math{1^{k_1}2^{k_2}\cdots{}n^{k_n},}
where \Math{k_i} is the number of cycles of the length \Math{i} in the permutation.
The \emph{characteristic polynomial} of permutation matrix \eqref{permrepEn} can
be written immediately from the cycle type of the corresponding permutation \Math{\pi(g)}:
\begin{equation}
\chi_{\rho(g)}\vect{\lambda}=\det\vect{\rho(g)-\lambda\idmat}
=\vect{\lambda-1}^{k_1}\vect{\lambda^2-1}^{k_2}\cdots\vect{\lambda^n-1}^{k_n}.
\label{charpolEn}
\end{equation}
The matrix form of permutations \eqref{regpermEn} representing the
 \emph{regular} action
\begin{equation}
    \Pi(g)\rightarrow{}\regrep(g)=
\Mone{\regrep(g)_{ij}},~~ \regrep(g)_{ij}=\delta_{e_ig,e_j},~~ i,j=1,\ldots,\wGN
    \label{regrepEn}
\end{equation}
is called the \emph{regular representation} --- this is a special case of \eqref{permrepEn}.
\par
For the sake of freedom of algebraic manipulations, one assumes usually that
\Math{\aNumbers} is an algebraically closed field --- a standard choice is
the field of complex numbers \Math{\C}.
If \Math{\aNumbers} is a field, then the set \Math{\Omega}
can be treated as a basis of linear vector space
\Math{\Hspace=\mathrm{Span}\vect{\omega_1,\cdots,\omega_\wSN}}.
\par
The field \Math{\C} is excessively large --- most of its elements are non-constructive.
What is really needed can be constructed as follows.
As is clear from \eqref{charpolEn}, all \emph{eigenvalues} of permutation matrices are
\Math{\ExpG}th roots
of unity, where \Math{\ExpG} is the \emph{exponent} of the group \Math{\wG}
--- the least common multiple of orders of the group elements.
The \Math{\ExpG}th roots of unity can be expressed in terms of \Math{\period}th
roots, where \Math{\period} is some divisor of \Math{\ExpG} called \emph{conductor}.
As a first step, we combine the roots of unity with \emph{natural numbers}
\Math{\N=\set{0,1,\ldots}} to construct the set
\Math{\aNumbers_{\period}=\N\ordset{\runi{}}} of polynomials
of the form \Math{n_1+n_2\runi{}+\cdots+n_{\period}\runi{}^{\period-1}},
where \Math{n_k\in\N}; \Math{\runi{}}
is \emph{primitive} \Math{\period}th root of unity, i.e. period of \Math{\runi{}} is
equal exactly to \Math{\period}.
For intuitive perception one could bear in mind the symbolics
\Math{\runi{}=\e^{2\pi{}i/\period}} for the primitive root,
but we will never use this representation.
The following \emph{algebraic} definitions are sufficient for all computations
\begin{enumerate}
    \item \emph{Multiplication}: \Math{\runi{}^k\times\runi{}^m=\runi{}^{k+m\mod{}\period}},
    \item \emph{Complex conjugation}: \Math{\cconj{\runi{}^k}=\runi{}^{\period-k}}.
\end{enumerate}
\par
If \Math{\period=1}, then
\Math{\aNumbers_1} is the \emph{semi-ring of natural numbers} \Math{\N}.
\par
If \Math{\period\geq2}, then \emph{negative integer numbers}  can be introduced via the definition
\Math{\vect{-1}=\sum\limits_{k=1}^{p-1}\runi{}^{\frac{\period}{p}k}\!\!,\text{\,where~}p
\text{~is any  factor of~}\period.} So we obtain the \emph{ring of integers} \Math{\Z}.
\par
If \Math{\period\geq3}, then the set
\Math{\aNumbers_\period} is a \emph{commutative ring} embeddable into
the field of complex numbers \Math{\C}. This is the ring of \emph{cyclotomic integers}:
\Math{\aNumbers_\period=\Z\ordset{\runi{}}/\braket{\Phi_\period\vect{\runi{}}}}.
Here \Math{\Phi_\period\vect{\runi{}}} is the \Math{\period}th \emph{cyclotomic polynomial}
--- the product of  the 
binomials
\Math{\runi{}-\zeta},
where \Math{\zeta} runs over \emph{all primitive} \Math{\period}th roots of unity.
\par
The ring \Math{\aNumbers_\period} is sufficient for almost all computations with finite
quantum models. For simplicity of linear algebra we extend the ring
\Math{\aNumbers_\period} to the \Math{\period}th \emph{cyclotomic field }
\Math{\Q_\period=\Q\ordset{\runi{}}/\braket{\Phi_\period\vect{\runi{}}}}.
When computing matrices of \emph{unitary} representations
\emph{square roots of dimensions} of representations arise as normalization factors.
Since square roots of integers are always cyclotomic integers we can treat all
irrationalities arising in computations
--- roots of unity and square roots of dimensions --- as belonging to a ring of
cyclotomic integers \Math{\aNumbers_n}  with some \Math{n} (usually \Math{n>\period}).
We can also construct a minimal \emph{abelian number field} \Math{\NF} containing a given
set of irrationalities. It is a subfield of the cyclotomic field \Math{\Q_n}.
The term \emph{abelian} means here that \Math{\NF} is an extension 
with abelian Galois group.
The command
\texttt{\textbf{Field(\textit{gens}\!)}} in the computer algebra system \GAP{}
\cite{gapEn} returns the \emph{smallest} field that contains all elements from
the list \textit{gens}. As to the finite quantum systems discussed in this paper,
the roots of unity and other irrationalities are only intermediate entities
in description of quantum behavior --- they disappear in the final ``observables''.
\par
Any linear representation of a finite group is equivalent to unitary,
since one can always construct invariant inner product from an arbitrary one by
``averaging over the group''.
Starting from, e.g.,  the
\emph{standard inner product} in \Math{\adimH}-dimensional Hilbert space \Math{\Hspace}
\begin{equation}
    \innerstandard{\phi}{\psi}\equiv\sum\limits_{i=1}^{\adimH}\cconj{\phi^i}\psi^i
\label{innerstdEn}
\end{equation}
we can come via the averaging to the \emph{invariant inner product}:
\begin{equation}
\inner{\phi}{\psi}\equiv\frac{\textstyle{1}}{\textstyle{\cabs{G}}}\sum\limits_{g\in{}G}
\!\innerstandard{U\!\vect{g}\phi}{U\vect{g}\psi}.
\label{innerinvEn}
\end{equation}
Here \Math{U} is a representation of a group \Math{G} in the space \Math{\Hspace}.
\par
An important transformation of group elements
--- an analog of change of coordinates in physics --- is the conjugation:
\Math{a^{-1}ga\rightarrow{}g',} \Math{g, g'\in\wG,} \Math{a\in\Aut{\wG}}.
Conjugation by an element of the group itself, i.e., if \Math{a\in\wG},
is called an \emph{inner automorphism}.
The equivalence classes with respect to the inner automorphisms are called
\emph{conjugacy classes}. The starting point in study of representations
of a group is its decomposition into conjugacy classes
\Mathh{\wG=\class{1}\sqcup\class{2}\sqcup\cdots\sqcup\class{\classN}.}
\par
The group multiplication induces \emph{multiplication} on the classes.
The product of \Math{\class{i}} and \Math{\class{j}} is
the \emph{multiset} of all possible products
\Math{ab,~a\in\class{i},~b\in\class{j},} decomposed into classes.
This multiplication is obviously commutative, since \Math{ab} and \Math{ba} belong to
the same class: \Math{ab\sim{}a^{-1}\vect{ab}a=ba}. Thus, the multiplication
table for classes is given by
\begin{equation}
    \class{i}\class{j}= \class{j}\class{i} = \sum\limits_{k=1}^{\classN}c_{ijk}\class{k}.
\label{classtabEn}
\end{equation}
The \emph{natural integers} \Math{c_{ijk}}
--- multiplicities of classes in the multisets --- are called
\emph{class coefficients}.
\par
This is a short list of main properties of linear representations of finite groups:
\begin{enumerate}
    \item
Any irreducible representation is contained in the regular
representation.
More specifically,  there exists 
matrix \Math{\transmatr} transforming
simultaneously all matrices
\eqref{regrepEn} to the form
\begin{equation}
    \transmatr^{-1}\regrep(g)\transmatr=
    \bmat
            \repirr_1(g) &&&
    \\[5pt]
    &
    \hspace*{-27pt}d_2\left\{
    \begin{matrix}
    \repirr_2(g)&&\\
    &\hspace*{-10pt}\ddots&\\
    &&\hspace*{-7pt}\repirr_2(g)
    \end{matrix}
    \right.
    &&
    \\
    &&\hspace*{-10pt}\ddots&\\
    &&&
    \hspace*{-25pt}d_{\classN}\left\{
    \begin{matrix}
    \repirr_{\classN}(g)&&\\
    &\hspace*{-10pt}\ddots&\\
    &&\hspace*{-7pt}\repirr_{\classN}(g)
    \end{matrix}
    \right.
    \emat,
\label{regrepdecompEn}
\end{equation}
and any irreducible representation is one of \Math{\repirr_j}'s.
The numbers of non-equivalent irreducible representation  and
conjugacy classes coincide. The number \Math{d_j} is the dimension
of the irreducible component \Math{\repirr_j} and simultaneously
the multiplicity of its occurrence in the regular representation.
It is clear from \eqref{regrepdecompEn} that for the dimensions of
irreducible representations the following relation holds:
\Math{d^2_1+d^2_2+\cdots+d^2_{\classN}=\cabs{\wG}=\wGN.} The
dimensions of irreducible representations divide the group order:
\Math{d_j\mid\wGN.}
    \item
Any irreducible representation \Math{\repirr_j} is determined uniquely
 by its \emph{character} \Math{\chi_j} defined as the trace of the representation
 matrix:
\Math{\chi_j\vect{g}=\mathrm{Tr}\repirr_j\vect{g}}.
This is a function on the conjugacy classes since
\Math{\chi_j\vect{g}=\chi_j\vect{a^{-1}ga}}.
Obviously, \Math{\chi_j\vect{\id}=d_j}.
    \item
A compact form of recording all irreducible representations is the
\emph{character table}.
The columns of this table are numbered by the conjugacy classes,
while its rows contain values of characters of non-equivalent representation:
\begin{center}
\begin{tabular}{c|cccc}
&\Math{\class{1}}&\Math{\class{2}}&\Math{\cdots}&\Math{\class{\classN}}\\\hline
\Math{\chi_1}&1&1&\Math{\cdots}&1\\
\Math{\chi_2}&\Math{\chi_2\vect{\class{1}}=d_2}&\Math{\chi_2\vect{\class{2}}}&
\Math{\cdots}&\Math{\chi_2\vect{\class{\classN}}}\\
\Math{\vdots}&\Math{\vdots}&\Math{\vdots}&&\Math{\vdots}\\
\Math{\chi_{\classN}}&\Math{\chi_{\classN}\vect{\class{1}}=d_{\classN}}&
\Math{\chi_{\classN}\vect{\class{2}}}&\Math{\cdots}&\Math{\chi_{\classN}\vect{\class{\classN}}}
\end{tabular}.
\end{center}
By convention, the 1st column corresponds to the identity class,
and the 1st row contains the \emph{trivial} representation.
\end{enumerate}
\section{Finite Quantum Systems}
In quantum mechanics all possible states of every physical system
are represented by vectors \Math{\psi} in a Hilbert space \Math{\Hspace}.
It is assumed that vectors \Math{\psi} and \Math{\psi'}
describe identical states if they are proportional through a complex factor:
\Math{\psi'=\lambda\psi,~\lambda\in\C}.
Evolution of the system from any initial state \Math{\psi_0} into
the corresponding final state \Math{\psi_T} is described by an \emph{unitary}
operator \Math{U}: \Math{\barket{\psi_T}=U\barket{\psi_0}}.
The unitarity means that \Math{U} belongs to the automorphism group
of the Hilbert space: \Math{U\in\Aut{\Hspace}}.
One may regard \Math{\Aut{\Hspace}} as a faithful representation of
respective abstract group \Math{\wG}.
In the continuous time the dynamics can be expressed by
the Schr\"{o}dinger equation
\Mathh{i\frac{\mathrm{d}}{\mathrm{d}t}\barket{\psi}=H\barket{\psi}}
in terms of the local \emph{Hermitian} operator \Math{H}
called the \emph{Hamiltonian} or \emph{energy operator}.
If  \Math{H} is independent of time,
then the relation \Math{U=\e^{-iHT}} holds.
\par
A finite quantum system is formulated in exactly the same way.
The only difference is that now the group \Math{\wG} is  a finite group
of order \Math{\wGN} having unitary representation \Math{\repq}
in \Math{\adimH}-dimensional Hilbert space \Math{\Hspace_{\adimH}} over some
abelian number field \Math{\NF} instead of \Math{\C}.
All possible evolution operators form the finite set
\Math{\set{U_1,\ldots,U_{\wGN}}} of unitary matrices from \Math{\repq}.
\par
Since the matrices \Math{U_j} are non-singular, one can always introduce
Hamiltonians by the formula
\Math{H_j=i\ln{}U_j\equiv\sum\limits_{k=0}^{p-1}\lambda_k{}U_j^k}, where \Math{p} is
period of \Math{U_j}, \Math{\lambda_k}'s are some coefficients%
\footnote{Note that the logarithmic function being essentially a
construction from continuous mathematics introduces into the
\Math{\lambda_k}'s a \emph{non-algebraic} element --- namely,
\Math{\pi}
---
expressed by \emph{infinite} sum of elements from \Math{\NF}.
In other words, the \Math{\lambda_k}'s are elements of a \emph{transcendental extension}
of  \Math{\NF}.}; but there is no need to do so.
\par
More generally, \emph{hermitian operators} \Math{A} describing \emph{observables}
in quantum formalism can be written as elements of the group algebra representation:
\Mathh{A=\sum\limits_{k=1}^{\wGN}\alpha_k{}U_k.}
\par
Finite groups --- unless they are many-component direct products
--- can be often generated by a small number of elements.
For example, all simple and all symmetric groups are generated by two elements.
The algorithm restoring the whole group from \Math{n_{g}} generators
is very simple. It is reduced to \Math{n_{g}\vect{\wGN-n_{g}-1}} group multiplications.
So the finite quantum models are well suited for
computer algebra methods.
\subsection{Reducing Quantum Dynamics to Permutations}
It follows from decomposition \eqref{regrepdecompEn}
that any \Math{\adimH}-dimensional representation \Math{\mathrm{U}}
can be extended to an \Math{\wSN}-dimensional representation
\Math{\mathrm{\widetilde{U}}} in a Hilbert space \Math{\Hspace_{\wSN}},
in such a way that the representation \Math{\mathrm{\widetilde{U}}}
corresponds to the \emph{permutation action} of the group \Math{\wG} on
some \Math{\wSN}-element set of entities \Math{\wS=\set{\ws_1,\ldots,\ws_{\wSN}}}.
It is clear that \Math{\wSN\geq\adimH}.
\par
The case when \Math{\wSN} is strictly greater than \Math{\adimH}
is most interesting.
Clearly, the additional ``hidden parameters''
 --- appearing in this case due to increase of the number of states (dimension of space)
 --- in no way can affect the data relating to the 
space \Math{\Hspace_{\adimH}} since both \Math{\Hspace_{\adimH}} and
its complement in \Math{\Hspace_{\wSN}}
are invariant subspaces of the extended space \Math{\Hspace_{\wSN}}.
Thus, \emph{any quantum problem} in \Math{\adimH}-dimensional
Hilbert space can be reformulated in terms of permutations of
\Math{\wSN} things.
\par
From the algorithmic point of view, manipulations with permutations are much more
efficient than the linear algebra operations with matrices. Of course,
degrees of permutations \Math{\wSN} might be much larger than dimensions of matrices
 \Math{\adimH}. However, the very possibility to
 \emph{reduce quantum dynamics to permutations}
 is much more important conceptually than the algorithmic issues.
\subsection{Connection with Observation. The Born Rule}
In quantum mechanics, the link between mathematical description and experiment is provided
by the \emph{Born rule}, stating that the \emph{probability}
to observe a quantum system being in the state \Math{\psi} by apparatus tuned to the state
\Math{\phi} is expressed by the number
\begin{equation}
\ProbBorn{\phi}{\psi} = \frac{\textstyle{\cabs{\inner{\phi}{\psi}}^2}}
{\textstyle{\inner{\phi}{\phi}\inner{\psi}{\psi}}}.
\label{BornEn}
\end{equation}
This expression can be rewritten in a form including the pair ``system--apparatus''
in more symmetric way
\Mathh{\ProbBorn{\phi}{\psi} =
\frac{\textstyle{\cabs{\inner{\phi}{\psi}}^2}}
{\textstyle{\cabs{\inner{\phi}{\psi}}^2+\Grassnorm{\phi\wedge\psi}^2}}.}
Here \Math{\phi\wedge\psi} is exterior (Grassmann) product of the vectors
\Math{\phi} and \Math{\psi}, which is the
\Math{\adimH(\adimH-1)/2}-dimensional vector with the components in the unitary basis
\Math{\vect{\phi\wedge\psi}^{ij}=\phi^i\psi^j-\phi^j\psi^i}
and with the square of norm
\Mathh{\Grassnorm{\phi\wedge\psi}^2=\sum\limits_{i=1}^{\adimH-1}
\sum\limits_{j=i}^{\adimH}\cabs{{\phi^{i}}\psi^{j}-{\phi^{j}}\psi^{i}}^2.}
\par
There are many philosophical speculations concerning the concept of probability and its
interpretation. However, what is really used in practice is the
\emph{frequency interpretation}: the probability is the ratio of the number of favorable
cases to the total number of cases.
In the case of finite sets there are no complications at all: the probability
is the rational number --- the ratio of the number of singled out elements of a set
to the total number of elements of the set.
\par
It can be shown that if data about states of a system and apparatus
are represented in the permutation basis by \emph{natural numbers}, then formula
 \eqref{BornEn} gives \emph{rational numbers} in the invariant subspaces of
 the permutation representation also, in spite of
 possible presence of cyclotomics and square roots in the intermediate computations.
\par
Let us consider permutation action of the group
\Math{\wG=\set{\wg_1,\ldots,\wg_{\wGN}}} on the set entities
\Math{\wS=\set{\ws_1,\ldots,\ws_{\wSN}}}.
We will describe the (quantum) states of the system and apparatus in the permutation
representation by the vectors
\begin{equation}
    \barket{n} = \Vthree{n_1}{\vdots}{n_{\wSN}} \text{~and~}
    \barket{m} = \Vthree{m_1}{\vdots}{m_{\wSN}},
    \label{natamplEn}
\end{equation}
respectively.
It is natural to assume that \Math{n_i} and \Math{m_i} are \emph{natural numbers},
interpreting them as the ``multiplicities of occurrences'' of the element
\Math{\ws_i} in the system and apparatus states, respectively.
In other words, the vectors \Math{\barket{n}} and \Math{\barket{m}} are elements
of \Math{\wSN}-dimensional module  \Math{\natmod_\wSN} over the semi-ring \Math{\N}.
Permutation action of \Math{\wG} on \Math{\wS} is equivalent to matrix
representation of \Math{\wG} in the module  \Math{\natmod_\wSN}.
We can turn the module  \Math{\natmod_\wSN} into the Hilbert space \Math{\Hspace_\wSN}
by extending the semi-ring  \Math{\N} to an abelian number field \Math{\NF} compatible with
the structure of 
\Math{\wG}.
\par
Of course, due to the symmetry the numbers \Math{n_i} and \Math{m_i} are not observable.
Only their \emph{invariant combinations} are observable.
Since the standard inner product defined in \eqref{innerstdEn}
is invariant for the permutation representation, in accordance with the Born rule
we have
\begin{equation}
    \ProbBorn{m}{n}=\frac{\vect{\sum_i{m_i}n_i}^2}{\sum_i{m_i}^2\sum_i{n_i}^2}.
\label{probpEn}
\end{equation}
It is clear that for non-vanishing natural vectors
 \Math{\barket{n}} and \Math{\barket{m}}  expression
\eqref{probpEn} is a rational number strictly greater than zero.
This means, in particular, that it is impossible to observe destructive quantum
interference here.
However, the \emph{destructive interference} of the vectors with natural components
can be observed in the proper invariant subspaces of the permutation representation.
\section{Example: Group of Permutations of Three Things}
\Math{\SymG{3}} is the smallest non-commutative group providing a non-trivial quantum behavior.
Nevertheless, \Math{\SymG{3}} has important applications in the lepton sector
of flavor physics.
The group consists of six elements
having the following representation by permutations
\begin{equation}
\wg_1=\vect{}\!,~\wg_2=\vect{2,3}\!,~\wg_3=\vect{1,3}\!,~\wg_4=\vect{1,2}\!,
    ~\wg_5=\vect{1,2,3}\!,~\wg_6=\vect{1,3,2}.
\label{S3elemsEn}
\end{equation}
The group can be generated by many pairs of its elements.
Let us choose, for instance, \Math{\wg_2} and \Math{\wg_6} as generators.
\Math{\SymG{3}} decomposes into three conjugacy classes
\begin{equation}
    \class{1}=\set{\wg_1},~~\class{2}=\set{\wg_2,~\wg_3,~\wg_4},
    ~~\class{3}=\set{\wg_5,~\wg_6}
    \label{S3classesEn}
\end{equation}
with the following multiplication table
\Mathh{\class{1}\class{j}=\class{j},~~\class{2}^2=3\class{1}+3\class{3},
~~\class{2}\class{3}=2\class{2},~~\class{3}^2=2\class{1}+\class{3}.}
The group \Math{\SymG{3}} has the following character table
\begin{equation}
    \text{\begin{tabular}{c|crr}
    &\Math{\class{1}}&\Math{\class{2}}&\Math{\class{3}}\\\hline
    \Math{\chi_1}&1&1&1\\
    \Math{\chi_2}&1&-1&1\\
    \Math{\chi_3}&2&0&-1
    \end{tabular}\enspace.}\label{S3tabEn}
\end{equation}
Matrices of permutation representation of generators are
\begin{equation}
    P_2=\Mthree{1}{~\cdot}{~\cdot}{\cdot}{~\cdot}{~1}{\cdot}{~1}{~\cdot}
    \text{~and}
    ~P_6=\Mthree{\cdot}{~\cdot}{~1}{1}{~\cdot}{~\cdot}{\cdot}{~1}{~\cdot}.
    \label{S3pmatsEn}
\end{equation}
The eigenvalues of \Math{P_2} and \Math{P_6} are \Math{\vect{1, 1, -1}} and
\Math{\vect{1, \runi{}, \runi{}^2}}, respectively; \Math{\runi{}}
is a primitive third root of unity with cyclotomic polynomial
\Math{\Phi_3\vect{\runi{}} = 1+\runi{}+\runi{}^2}.
\par
Since ~any permutation representation contains one-dimensional invariant subspace with
the basis vector \Math{\vect{1,\ldots,1}^\mathrm{T}}, the only possible structure
of decomposition of permutation representation into irreducible parts is the following
\begin{equation}
    \widetilde{U}_j=\Mtwo{1}{0}{0}{U_j},~~j =1,\ldots,6,
    \label{S3permqEn}
\end{equation}
where the matrices 1 and \Math{U_j} correspond to
one-dimensional trivial (character \Math{\chi_1})
and two-dimensional faithful (character \Math{\chi_3}) representations, respectively.
\par
To construct decomposition \eqref{S3permqEn} we should determine matrices \Math{U_j}
and \Math{\transmatr} such that \Math{\widetilde{U}_j=\transmatr^{-1}P_j\transmatr}.
In addition we impose unitarity on all the matrices.
Clearly, it suffices to perform the procedure only for matrices of generators.
There are different ways to construct decomposition \eqref{S3permqEn}.
\par
If we start with the diagonalization of \Math{P_6}, we come to the following%
\footnote{Note the peculiarity of representation \eqref{S3umatsEn}
--- its matrices are very similar to matrices of permutations: there is exactly
one non-zero entry in each column and in each row. But in contrast to permutation
matrices in which any non-zero entry is \emph{unity},
non-zeros in \eqref{S3umatsEn} are \emph{roots of unity}.
This is because \Math{\SymG{3}} is one of the so-called
\emph{monomial groups} \cite{Kirillov} for which all irreducible representations can be
 constructed as induced from one-dimensional representations of their subgroups
--- choosing diagonal form for \Math{U_6} is just equivalent to
inducing \eqref{S3umatsEn} from representation of cyclic subgroup
\Math{\CyclG{3}\leq\SymG{3}}.
Most groups, at least of small orders,
are just monomial. For example, it can be checked with the help of \GAP{} that the total number of all non-isomorphic groups
of order \Math{<384} is equal to 67424, but only 249 of them are \emph{non-monomial}.
The minimal non-monomial group is the 24-element group \Math{\SL{2}{3}} of \Math{2\times2}
matrices in the characteristic 3 with unit determinants.}
\begin{align}
        U_1=\Mtwo{1}{0}{0}{1},~U_2=\Mtwo{0}{\runi{}^2}{\runi{}}{0},
        ~U_3=\Mtwo{0}{\runi{}}{\runi{}^2}{0},\nonumber\\[-6pt]
        \label{S3umatsEn}\\[-6pt]
        ~U_4=\Mtwo{0}{1}{1}{0},
        ~U_5=\Mtwo{\runi{}^2}{0}{0}{\runi{}},~U_6=\Mtwo{\runi{}}{0}{0}{\runi{}^2}.\nonumber
    \end{align}
The transformation matrix (up to inessential degrees of freedom for its
entries) takes the following form
\begin{equation}
\transmatr=\frac{1}{\sqrt{3}}
    \Mthree{1}{1}{\runisymb^2}
     {1}{\runisymb^2}{1}
     {1}{\runisymb}{\runisymb},~~~~
\transmatr^{-1}=\frac{1}{\sqrt{3}}
    \Mthree{1}{1}{1}
     {1}{\runisymb}{\runisymb^2}
     {\runisymb}{1}{\runisymb^2}.
\label{transS3monomial}
\end{equation}
\par
Otherwise, the diagonalization of \Math{P_2} leads to another second component
of decomposition \eqref{S3permqEn} (we present here only the generator matrices)
\Mathh{
U'_2=\Mtwo{1}{0}{0}{-1},
~~~~U'_6=\Mtwo{-\frac{1}{2}}{\frac{\sqrt{3}}{2}}{-\frac{\sqrt{3}}{2}}{-\frac{1}{2}}.
}
The transformation matrix in this case takes the form
\begin{equation}
\transmatrprim
=\Mthree{\frac{1}{\sqrt{3}}}{~\,\,\sqrt{\frac{2}{3}}}{~0}
          {\frac{1}{\sqrt{3}}}{\,-\!\frac{1}{\sqrt{6}}}{-\!\frac{1}{\sqrt{2}}}
          {\frac{1}{\sqrt{3}}}{\,-\!\frac{1}{\sqrt{6}}}{~\,\,\frac{1}{\sqrt{2}}},
~~~~
\transmatrprim^{-1}
=\Mthree{\frac{1}{\sqrt{3}}}{~\,\frac{1}{\sqrt{3}}}{~\frac{1}{\sqrt{3}}}
        {\sqrt{\frac{2}{3}}}{\,-\!\frac{1}{\sqrt{6}}}{-\!\frac{1}{\sqrt{6}}}
        {~0}{\,-\!\frac{1}{\sqrt{2}}}{~\,\,\frac{1}{\sqrt{2}}}.
\label{transS3sqrt}
\end{equation}
The matrix \Math{\transmatrprim} is known in particle physics under the names
\emph{Harrison-Perkins-Scott}  or \emph{tribimaximal} mixing matrix.
It is used to description of neutrino oscillation data.
\par
The information about ``quantum behavior'' is encoded, in fact,
in transformation matrices like \eqref{transS3monomial} or \eqref{transS3sqrt}.
\par
Let \Math{\barket{n} = \Vthree{n_1}{n_2}{n_3}} and
\Math{\barket{m} = \Vthree{m_1}{m_2}{m_3}} be system and apparatus
state vectors in the ``permutation'' basis. Transformation of
these vectors from the permutation to ``quantum'' basis with the
help of, say, \eqref{transS3monomial} leades to
\begin{align*}
    \barket{\widetilde{\psi}}=\transmatr^{-1}\barket{n}
    &=\frac{1}{\sqrt{3}}\Vthree{n_1+n_2+n_3}
    {n_1+n_2\runisymb+n_3\runisymb^2}{n_1\runisymb+n_2+n_3\runisymb^2},\\
    \barket{\widetilde{\phi}}=\transmatr^{-1}\barket{m}
    &=\frac{1}{\sqrt{3}}\Vthree{m_1+m_2+m_3}
    {m_1+m_2\runisymb+m_3\runisymb^2}{m_1\runisymb+m_2+m_3\runisymb^2}.
\end{align*}
Projections of the vectors onto two-dimensional invariant subspace
 are:
\begin{equation}
    \barket{\psi} = \frac{1}{\sqrt{3}}\Vtwo{n_1+n_2\runisymb+n_3\runisymb^2}
    {n_1\runisymb+n_2+n_3\runisymb^2},~~~~
    \barket{\phi} = \frac{1}{\sqrt{3}}\Vtwo{m_1+m_2\runisymb+m_3\runisymb^2}
    {m_1\runisymb+m_2+m_3\runisymb^2}.
    \label{promonom}
\end{equation}
The same manipulation with matrix \eqref{transS3sqrt}   leads to
\begin{equation}
    \barket{\psi'} = \Vtwo{n_1\sqrt{\frac{2}{3}}-n_2\frac{1}{\sqrt{6}}-n_3\frac{1}{\sqrt{6}}}
    {-n_2\frac{1}{\sqrt{2}}+n_3\frac{1}{\sqrt{2}}},~~~~
    \barket{\phi'} = \Vtwo{m_1\sqrt{\frac{2}{3}}-m_2\frac{1}{\sqrt{6}}-m_3\frac{1}{\sqrt{6}}}
    {-m_2\frac{1}{\sqrt{2}}+m_3\frac{1}{\sqrt{2}}}.
    \label{prosquare}
\end{equation}
\par
Constituents of Born's probability \eqref{BornEn} for the two-dimensional subsystem
 --- clearly, the same in both cases \eqref{promonom} and \eqref{prosquare} --- are
\begin{equation}
    \inner{\psi\!}{\!\psi}=\invarQ{3}{n}{n}-\frac{1}{3}\invarL{3}{n}^2,
\label{Born2den1En}
\end{equation}
\begin{equation}
    \inner{\phi\!}{\!\phi}=\invarQ{3}{m}{m}-\frac{1}{3}\invarL{3}{m}^2,
\label{Born2den2En}
\end{equation}
\begin{equation}
\cabs{\inner{\phi\!}{\!\psi}}^2=
\vect{\invarQ{3}{m}{n}-\frac{1}{3}\invarL{3}{m}\invarL{3}{n}}^2,
\label{Born2numEn}
\end{equation}
where \Math{\invarL{\wSN}{n}=\sum\limits_{i=1}^{\wSN}n_i} and
\Math{\invarQ{\wSN}{m}{n}=\sum\limits_{i=1}^{\wSN}m_in_i} are linear and
quadratic permutation invariants, respectively.
\par
Note that:
\begin{enumerate}
    \item Expressions \eqref{Born2den1En}--\eqref{Born2numEn} consist of the
    \emph{invariants of permutation representation}.
    This is a manifestation of fundamental role of permutations in quantum description.
    \item
Expressions \eqref{Born2den1En} and \eqref{Born2den2En} are always
positive rational numbers for \Math{\barket{n}} and
\Math{\barket{m}} with different components.
    \item
Conditions for \emph{destructive quantum interference} ---
vanishing Born's probability ---    are determined by the equation
\Mathh{3\vect{m_1n_1+m_2n_2+m_3n_3}-\vect{m_1+m_2+m_3}\vect{n_1+n_2+n_3}=0.}
This equation has infinitely many solutions in natural numbers.
An example of such a solution is:
\Math{{\barket{n} = \Vthree{1}{1}{2},~~\barket{m} = \Vthree{1}{3}{2}}}.
\end{enumerate}
Thus, we have obtained
essential features of quantum behavior from ``permutation dynamics''
and ``natural'' interpretation \eqref{natamplEn} of quantum amplitude
by a simple transition to invariant subspaces.
\par
Recall once more that any permutation representation contains the
trivial one-dimensional subrepresentation and, hence, has
\Math{\vect{\wSN-1}}-dimensional invariant subspace. The inner
product in this subspace can be expressed in terms of the
permutation invariants by the formula
\Mathh{\inner{\phi\!}{\!\psi}=
\invarQ{\wSN}{m}{n}-\frac{1}{\wSN}\invarL{\wSN}{m}\invarL{\wSN}{n}.}
The identity
\Math{\displaystyle\invarQ{\wSN}{n}{n}-\frac{1}{\wSN}\invarL{\wSN}{n}^2\equiv
\frac{1}{\wSN^2}\sum\limits_{i=1}^\wSN\vect{\invarL{\wSN}{n}-\wSN{}n_i}^2}
shows explicitly that \Math{\inner{\psi\!}{\!\psi}>0} for
\Math{\barket{n}} with different components \Math{n_i}. This inner
product does not contain irrationalities for natural
\Math{\barket{n}} and \Math{\barket{m}}. This is not the case for
other invariant subspaces. Nevertheless irrationalities disappear
in the squared modulus of the inner product
\Math{\cabs{\inner{\phi\!}{\!\psi}}^2}. To give a simple
illustration let us consider the cyclic group \Math{\CyclG{3}}.
Its three-dimensional permutation representation decomposes into
three one-dimensional irreducible components. E.g., for the
generator \Math{g=\vect{1,2,3}} of  \Math{\CyclG{3}} we have
\Mathh{P=\Mthree{\,\cdot}{\,1}{\,\cdot}{\,\cdot}{\,\cdot}{\,1}{\,1}{\,\cdot}{\,\cdot}
\longrightarrow
\widetilde{U}=\Mthree{\,1}{\,0}{\,0}{\,0}{\,\runi{}}{\,0}{\,0}{\,0}{\,\runi{}^2},~~
{\runi{}}~ \text{is a primitive third root of unity}.} The inner
product in one-dimensional subspace corresponding to the
eigenvalue, say \Math{\runi{}}, contains irrationalities:
\Math{\displaystyle\inner{\phi\!}{\!\psi}=\frac{1}{3}\vect{\invarQ{3}{m}{n}
+\runi{}C(m,n)+\runi{}^2C'(m,n)}}, but
\Math{\displaystyle\cabs{\inner{\phi\!}{\!\psi}}^2
=\frac{1}{9}\vect{\invarQ{3}{m}{m}\!-C(m,m)}\vect{\invarQ{3}{n}{n}\!-C(n,n)}}
is free of them. The invariants \Math{C(m,n)=m_1n_3+m_2n_1+m_3n_2}
and \Math{C'(m,n)=m_1n_2+m_2n_3+m_3n_1} are specific for the group
\Math{\CyclG{3}} in contrast to \Math{\invarL{\wSN}{n}} and
\Math{\invarQ{\wSN}{m}{n}} that are common to all permutation
groups.
\section{Finite Symmetry Groups in Particle Physics}
At present, all observations concerning fundamental particles \cite{RevPartPhys}
are compatible with the Standard Model (SM).
The SM is a gauge theory with the group of internal (gauge) symmetries
\Math{\iG=\SU{3}\times\SU{2}\times\U{1}}. In the context of Grand Unified Theory
(GUT) \Math{\iG} is assumed to be a subgroup of some larger (simple) group.
With respect to space-time symmetries, the elementary particles are divided into
two classes:
\emph{bosons}, responsible for physical forces (roughly speaking,
they are elements of the gauge group) and \emph{fermions},
usually treated as particles of matter. The fermions of the SM are divided into
three \emph{generations} of \emph{quarks} and \emph{leptons} as follows
(antiparticles are omitted for brevity):\\[-15pt]
\begin{center}
\begin{tabular}{c|c|c|c}
\multicolumn{4}{c}{\hspace{90pt}Generations}
\\
&1&2&3
\\\hline
\begin{tabular}{l}
Up-quarks
\\
Down-quarks~~~
\end{tabular}
&
\begin{tabular}{l}
Up\hspace{67pt}\Math{u~~}
\\
Down\hspace{55pt}\Math{d~~}
\end{tabular}
&
\begin{tabular}{l}
Charm\hspace{40pt}\Math{c~~}
\\
Strange\hspace{37pt}\Math{s~~}
\end{tabular}
&
\begin{tabular}{l}
Top\hspace{40pt}\Math{t~~}
\\
Bottom\hspace{25pt}\Math{b~~}
\end{tabular}
\\\hline
\begin{tabular}{l}
Charged leptons
\\
Neutrinos
\end{tabular}
&
\begin{tabular}{lc}
Electron&\Math{e^-}
\\
Electron neutrino&\Math{\nu_e}
\end{tabular}
&
\begin{tabular}{lc}
Muon&\Math{\mu^-}
\\
Muon neutrino&\Math{\nu_\mu}
\end{tabular}
&
\begin{tabular}{lc}
Tau&\Math{\tau^-}
\\
Tau neutrino&\Math{\nu_\tau}
\end{tabular}
\end{tabular}
\end{center}
Between generations particles  differ only by their mass  and quantum property
called \emph{flavor}.
The flavor changing transitions --- taking place in such phenomena as weak decays of quarks
and neutrino oscillations --- are described
by \Math{3\times3} unitary \emph{mixing matrices}.
The outputs of experiments allow to calculate magnitudes of elements of these matrices.
\par
In the case of quarks (``\emph{in the quark sector}''), the mixing
matrix describing transitions between up- and down-type quarks is
the \emph{Cabibbo--Kobayashi--Maskawa} (CKM) matrix \Mathh {
V_{\text{CKM}}=\Mthree{V_{ud}}{V_{us}}{V_{ub}}
       {V_{cd}}{V_{cs}}{V_{cb}}
       {V_{td}}{V_{ts}}{V_{tb}},
}
where \Math{\cabs{V_{\alpha\beta}}^2} represents the probability that the quark (of
 flavor)  \Math{\beta} decays into a quark \Math{\alpha}.
The current experimental data rounded to three significant digits  are:
\Mathh{
\Mthree{\cabs{V_{ud}}}{\cabs{V_{us}}}{\cabs{V_{ub}}}
       {\cabs{V_{cd}}}{\cabs{V_{cs}}}{\cabs{V_{cb}}}
       {\cabs{V_{td}}}{\cabs{V_{ts}}}{\cabs{V_{tb}}}
=\Mthree{0.974}{~0.225}{~0.004}
        {0.225}{~0.974}{~0.041}
        {0.009}{~0.040}{~0.999}.
}
More precise values can be found in \cite{RevPartPhys}.
\par
In the lepton sector weak interaction processes are described by the
\emph{ Pon\-te\-corvo--Maki--Nakagawa--Sakata} (PMNS) mixing matrix
\Mathh
{
U_{\text{PMNS}}=\Mthree{U_{e1}}{U_{e2}}{U_{e3}}
       {U_{\mu1}}{U_{\mu2}}{U_{\mu3}}
       {U_{\tau1}}{U_{\tau2}}{U_{\tau3}}.
}
Here indices \Math{e, \mu, \tau} correspond to neutrino flavors ---
this means that the neutrinos \Math{\nu_e, \nu_\mu, \nu_\tau} are produced with
\Math{e^+, \mu^+, \tau^+} (or produce \Math{e^-, \mu^-, \tau^-}),
respectively, in weak processes. The indices \Math{1, 2, 3} correspond to the
\emph{mass eigenstates}, i.e.,  neutrinos
 \Math{\nu_1, \nu_2, \nu_3} with definite masses  \Math{m_1, m_2, m_3}.
Numerous experiments with solar, atmospheric, reactor, and
accelerator neutrinos indicate the existence of discrete
symmetries that can not be deduced from the SM. The
phenomenological pattern is the following \cite{Smirnov}:
\begin{enumerate}
    \item  \Math{\nu_\mu} and
\Math{\nu_\tau} flavors are presented with equal weights in all three mass\\
eigenstates
\Math{\nu_1, \nu_2, \nu_3} (this is called ``\emph{bi-maximal mixing}''):\\
\Math{\cabs{U_{\mu{i}}}^2=\cabs{U_{\tau{i}}}^2},~~ \Math{i=1,2,3};
    \item
all three flavors are presented equally in \Math{\nu_2}
 (``\emph{trimaximal mixing}''):\\
\Math{\cabs{U_{e{2}}}^2=\cabs{U_{\mu{2}}}^2=\cabs{U_{\tau{2}}}^2};
    \item  \Math{\nu_e} is absent in \Math{\nu_3}: \Math{\cabs{U_{\mu3}}^2=0}.
\end{enumerate}
These relations together with the normalization condition for probabilities
allow to determine moduli-squared of all matrix elements:
\begin{equation}
    \Mone{\cabs{U_{l{i}}}^2}
    =\Mthree{~\fra{2}{3}}{~~\fra{1}{3}}{~~0~}
            {\\[-8pt]~\fra{1}{6}}{~~\fra{1}{3}}{~~\fra{1}{2}~}
            {\\[-8pt]~\fra{1}{6}}{~~\fra{1}{3}}{~~\fra{1}{2}~}.
\label{tribi2}
\end{equation}
A particular form of unitary matrix satisfying  data
\eqref{tribi2} was suggested by Harrison, Perkins, and Scott in
\cite{HPS02}:
\begin{equation}
    U_{\text{TB}}
    =\Mthree{~\,\sqrt{\fra{2}{3}}}{~~\fra{1}{\sqrt{3}}}{~~0}
            {-\fra{1}{\sqrt{6}}}{~~\fra{1}{\sqrt{3}}}{~-\!\fra{1}{\sqrt{2}}}
            {-\fra{1}{\sqrt{6}}}{~~\fra{1}{\sqrt{3}}}{~~~~\,\fra{1}{\sqrt{2}}}.
\label{tribi}
\end{equation}
This so-called  \emph{tribimaximal} (TB) mixing matrix coincides
--- up to the trivial permutation of two columns corresponding to the renaming
\Math{\nu_1\rightleftarrows\nu_2} of  states --- with
transformation matrix \eqref{transS3sqrt} decomposing the natural
permutation representation of the group \Math{\SymG{3}} into
irreducible components. This means that we can identify the flavor
basis with the representation basis of permutations of three
things, and the mass basis is a basis of irreducible decomposition
of this representation. In \cite{HS03} Harrison and Scott study in
detail connections of the neutrino mass matrix with the character
table and class algebra of the group \Math{\SymG{3}}. At present,
much effort is devoted to the construction and study of models
based on finite flavor symmetries (for recent reviews, see, for
example, \cite{Ishimori,Ludlgen}). The most popular groups for
constructing such models  are:
\begin{itemize}
    \item \Math{\mathsf{T}=\AltG{4}} --- the tetrahedral group;
    \item   \Math{\TprimG} --- the double covering of \Math{\AltG{4}};
    \item \Math{\mathsf{O}=\SymG{4}} --- the octahedral group;
    \item \Math{\mathsf{I}=\AltG{5}} --- the icosahedral group;
    \item \Math{\DihG{N}} --- the dihedral groups (\Math{N} even);
    \item \Math{\QuatG{N}} --- the quaternionic groups (4 divides \Math{N});
    \item   \Math{\Sigma\vect{2N^2}}
--- the groups in this series have the structure
\Math{\vect{\CyclG{N}\times\CyclG{N}}\rtimes\CyclG{2}};
    \item   \Math{\Delta\vect{3N^2}} --- the structure
\Math{\vect{\CyclG{N}\times\CyclG{N}}\rtimes\CyclG{3}};
    \item   \Math{\Sigma\vect{3N^3}}
--- the structure
\Math{\vect{\CyclG{N}\times\CyclG{N}\times\CyclG{N}}\rtimes\CyclG{3}};
    \item   \Math{\Delta\vect{6N^2}} --- the structure
\Math{\vect{\CyclG{N}\times\CyclG{N}}\rtimes\SymG{3}}.
\end{itemize}
\par
As to the quark sector, observations do not give such sharp picture as in the lepton case.
In \cite{BlumHagedorn} the \Math{\DihG{14}} symmetry was suggested for explanation of the
value of the Cabibbo angle (one of the parameters of the CKM matrix), but without
any connection with the leptonic symmetries.
The natural attempts to find discrete symmetries unifying leptons and quarks
still remain not very successful, though there are some encouraging observations,
for example, the \emph{quark-lepton complementarity} (QLC) ---
observation that the sum of quark and lepton mixing angles is equal approximately to
 \Math{\pi/4}.
\par
The origin of finite symmetries among fundamental particles is
unclear. There are different attempts to explain --- sometimes
looking a bit complicated and artificial, for example, these
symmetries are treated as symmetries of manifolds arising at
compactification of a higher dimensional theory to four spacetime
dimensions \cite{Altarelli}. The idea that symmeties at the most
fundamental level are \textit{per se} finite looks more attractive
in our opinion. In this approach, unitary groups used in physical
theories can be treated simply as repositories of all finite
groups having faithful representation of corresponding dimensions:
\Math{\U{n}}
contains all finite groups with faithful \Math{n}-dimensional
representations. 
Of course, due to redundancy of the field \Math{\C},
\Math{\U{n}} is not a minimal group with this property.
\par
Such small groups as \Math{\SymG{3},~\AltG{4}}, etc. are most
likely only remnants of large combinations of more fundamental
finite symmetries that are expected to exist at the GUT scale.
Unfortunately the GUT scale (\Math{10^{16}} GeV) being close to
the Planck scale (\Math{10^{19}} GeV) is out of reach of
experiments (the most powerful colliders to date can provide only
about \Math{10^{4}} GeV).
Thus, the only practical way is to construct 
models, study them by the computational group theory methods, and
compare consequences of these models with available experimental
data.
\section*{Conclusion}
``Finite'' analysis shows that quantum behavior
is a manifestation of indistinguishability of objects, i.e.,
fundamental impossibility to trace the identity of homogeneous
objects in the process of their evolution.
\par
Only ``statistical'' statements about numbers of certain
invariant combinations of elements may have objective significance.
These statements can be expressed in terms of group invariants and
natural numbers characterizing symmetry groups, such as dimensions of its
representations, class coefficients etc.
\par
Any quantum mechanical problem can be reduced to permutations
since permutation representations contain all other
representations. This --- together with natural interpretation of
quantum amplitudes as vectors of ``multiplicities of occurences''
of underlying permuted entities --- makes quantum mechanical
problems constructive and particularly suitable for their study by
computer algebra and computational group theory methods.
\par
The models based on finite groups are now extensively studied in particle physics,
since there are strong observational evidences of  finite symmetries
in fundamental physical processes.
\paragraph{Acknowledgment.}
The work was supported by the grants 01-01-00200 from the Russian Foundation for Basic
Research and 3810.2010.2 from the Ministry of Education and Science of
the Russian Federation.
\par

\end{document}